\documentclass[conference, 9 pt]{IEEEtran}
\IEEEoverridecommandlockouts

\usepackage{array}

\usepackage{amsmath, amssymb, amsthm} 

\usepackage{mathtools}

\usepackage{fixmath}

\usepackage{booktabs,multirow,threeparttable}
\usepackage{tabularx,ragged2e}
\newcolumntype{L}{>{\RaggedRight\arraybackslash}X}
\newcolumntype{C}{>{\centering\arraybackslash}X}
\usepackage{threeparttable}
\usepackage{standalone}
\usepackage{dcolumn}
\usepackage{hhline}
\usepackage{multienum}
\usepackage{multirow}
\usepackage{colortbl}
\usepackage{boldline}

\usepackage{algorithm}
\usepackage[noend]{algorithmic}

\usepackage{url}

\usepackage{flafter}

\usepackage[bottom,symbol]{footmisc}

\usepackage{todonotes}

\usepackage{listings}
\definecolor{light-gray}{gray}{0.95} 
\lstdefinestyle{BashInputStyle}{
  language=bash,
  basicstyle=\small\sffamily,
  numbers=left,
  numberstyle=\tiny,
  numbersep=3pt,
  frame=tb,
  columns=fullflexible,
  backgroundcolor=\color{yellow!20},
  linewidth=0.9\linewidth,
  xleftmargin=0.1\linewidth
}

\newcommand{\REVone}{\textcolor{black}}

\newcommand{\DACREVone}{\textcolor{black}}
\usepackage{balance}
\usepackage{subcaption}

\usepackage{ulem}
\usepackage{amsmath,amssymb,amsfonts}
\usepackage{algorithmic}
\usepackage{graphicx}
\usepackage{subcaption}
\usepackage{textcomp}
\usepackage{xcolor}
\usepackage{multirow}
\usepackage{amsmath}
\usepackage{comment}
\usepackage{cite}
\usepackage{enumitem}
\usepackage{todonotes}
\def\BibTeX{{\rm B\kern-.05em{\sc i\kern-.025em b}\kern-.08em
    T\kern-.1667em\lower.7ex\hbox{E}\kern-.125emX}}

\title{Algorithms and Hardware for Efficient Processing of Logic-based Neural Networks\thanks{ \scriptsize This research is supported in part by a grant from the Software and Hardware Foundations program of the National Science Foundation.}}

\author{\IEEEauthorblockN{Jingkai Hong*, Arash Fayyazi*\thanks{*Jingkai Hong and Arash Fayyazi contributed equally to this work.}, Amirhossein Esmaili, Mahdi Nazemi, and Massoud Pedram}
\IEEEauthorblockA{Department of Electrical \& Computer Engineering, University of Southern California, Los Angeles, CA, USA\\
    \url{{jingkaih,fayyazi,esmailid,nazemi,pedram}@usc.edu}}
}

\begin{document}
\maketitle
\thispagestyle{plain}
\pagestyle{plain}


\begin{abstract}
Recent efforts to improve the performance of neural network (NN) accelerators that meet today's application requirements have given rise to a new trend of logic-based NN inference relying on fixed-function combinational logic (FFCL). This paper presents an innovative optimization methodology for compiling and mapping NNs utilizing FFCL into a logic processor. The presented method maps FFCL blocks to a set of Boolean functions where Boolean operations in each function are mapped to high-performance, low-latency, parallelized processing elements. Graph partitioning and scheduling algorithms are presented to handle FFCL blocks that cannot straightforwardly fit the logic processor. Our experimental evaluations across several datasets and NNs demonstrate the superior performance of our framework in terms of the inference throughput compared to prior art NN accelerators. We achieve 25x higher throughput compared with the XNOR-based accelerator for VGG16 model that can be amplified 5x deploying the graph partitioning and merging algorithms.
\end{abstract}

\section{Introduction}


Deep neural networks (DNNs) provide state-of-the-art (SoA) performance in various artificial intelligence applications and have surpassed the accuracy of conventional machine learning models in many challenging domains, including computer vision
\cite{DBLP:conf/bmvc/ZagoruykoK16,DBLP:conf/cvpr/HuangLMW17} 
and natural language processing \cite{DBLP:journals/neco/HochreiterS97,DBLP:conf/naacl/DevlinCLT19}. The emergence of more complex DNN models such as \textit{transformers} \cite{DBLP:journals/corr/VaswaniSPUJGKP17} and \textit{MLPMixers} \cite{DBLP:conf/nips/TolstikhinHKBZU21} is a key reason for the remarkable performance of DNNs in many application domains but these models impose huge compute and memory resource requirements.

To improve the efficiency of DNNs, some prior work formulates the problem of efficient processing of neural networks as a Boolean logic minimization problem where ultimately, logic expressions compute the output of various filters/neurons. NullaNet \cite{DBLP:conf/aspdac/NazemiPP19} optimizes a target DNN for a given dataset and maps essential parts of the computation in DNNs to logic blocks, such as look-up tables (LUTs) on FPGAs. In most cases, including those used in this study, the average accuracy drop for binary implementation is less than 4\%. 

The issue is once the target fixed-function combinational logic (FFCL) for a specific NN model is synthesized, the synthesized fabric can only be used for the inference task of that given NN model\DACREVone{, which makes ASIC realization of FFCL impractical.} 
In addition, since neurons designed for SoA NNs include tens to hundreds of inputs, the obtained Boolean logic expression can be huge. \DACREVone{Our
experiments show that it is impossible to map some generated Boolean logic functions onto a single FPGA because LUTs are used up and cannot accommodate such huge Boolean functions.}
A logic processor comprising a Boolean logic unit, in which the Boolean logic unit is responsible for performing Boolean operations of each Boolean function associated with an FFCL block extracted from a binary neural network (BNN), as introduced in \cite{DBLP:conf/fccm/NazemiFEKSP21}, is critical to doing inference with different BNNs. Therefore, designing configurable efficient logic processors as logic-based inference engines, which can be used in a variety of applications with different DNN models, is highly desirable.

The compilation and scheduling of an arbitrary logic graph associated with a Boolean function to be mapped onto a logic processor is a demanding task from the viewpoint of the compiler design. The compiler needs to detect and group the operations of all gates that can be executed simultaneously, considering hardware resource limitations (i.e., the number of Boolean logic units per logic processor). 
\REVone{To address these challenges}, we design a many-core logic processor and introduce 
novel techniques for compiling and mapping BNNs that utilize FFCL into this logic processor. Furthermore, we present an original graph partitioning algorithm for handling very large logic graphs.
The contributions of the paper are as follows: 
\begin{itemize}
    \item We present the design of a  logic processor which can process large Boolean functions. 
    \item We present an innovative optimization methodology for compiling and mapping BNNs utilizing FFCL into this logic processor. \REVone{The proposed compiler generates customized instructions for static scheduling of all operations of the logic graph during inference.}
    \item We present a method to map FFCL blocks to a set of Boolean functions where Boolean operations in each function are mapped to high-performance, low-latency, and parallelized processing elements.
    \item We present a graph partitioning algorithm to efficiently deal with very large \DACREVone{Boolean} logic graphs.
    \item Our experimental evaluations across several datasets and NNs demonstrate the superior performance of our framework in terms of the inference throughput compared to prior art NN accelerators. We achieve 25x higher throughput compared with the XNOR-based accelerator for VGG16 model \cite{DBLP:journals/corr/SimonyanZ14a} that can be amplified 5x deploying the graph partitioning and merging algorithms.
    
\end{itemize}
While the scope of this work is focused on the algorithms for enabling the execution of arbitrary-size FFCL blocks on a logic processing fabric, we also present the results of using our algorithms and hardware design on an FPGA to determine their efficacy. The algorithms and hardware can be used for an ASIC design as well.

\section{Terminology and Notation} \label{sec:notation}


This section provides the terminology and notation used throughout this paper:
\begin{itemize}
    \item \textbf{Fixed-function combinational logic (FFCL) block:} A netlist of a combinational logic circuit written in a hardware description language, such as Verilog.
    \item \textbf{Logic processing element (LPE):} A \DACREVone{programmable} hardware block which performs (two-input) logic operations such as AND, OR, XOR, etc.
    \item \textbf{Logic processing vector (LPV):} A hardware block that contains a fixed number of LPEs. LPVs are linearly ordered relative to each other.
    \item \textbf{Logic processing unit (LPU):} A hardware block that contains a fixed number of LPVs, also called a \textbf{logic processor} in this paper.
    \item \textbf{Maximal feasible subgraph (MFG):} A directed acyclic graph (where nodes are Boolean operations and edges are data dependencies) \DACREVone{greedily extracted from an FFCL without exceeding the LPU's capacity when mapping to an LPU.}
    \item \textbf{Full path balancing (FPB):} Equalizing the logic depth of all propagation paths from circuit inputs \DACREVone{(i.e., primary inputs)} to circuit outputs \DACREVone{(i.e., primary outputs)}. It guarantees all input-output paths have the same number of gates on them. 

\end{itemize}


\section{Proposed Design Flow}\label{sec:Proposed_method}
\begin{figure} [tb!]
\vspace{-10pt}
    \centering
    \includegraphics[width=\columnwidth]{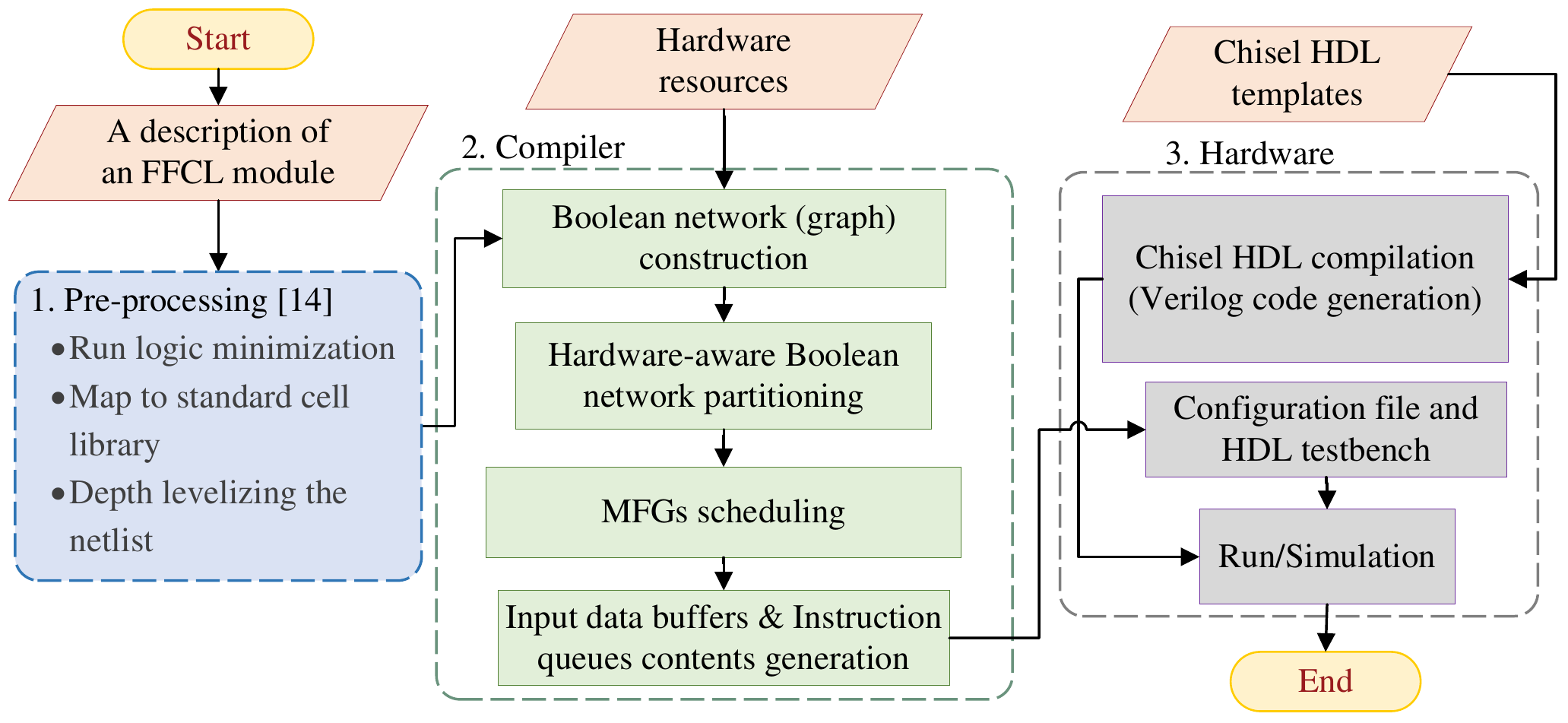}
    \caption{ \small An overview of the proposed framework.}
    \label{fig:flow}
\end{figure}

The overall flow of the proposed framework is as depicted in Fig. \ref{fig:flow}.
The input to the flow is a description of an FFCL block in the Verilog language. Please note that the framework can be structured to accept any specification of an FFCL block as the input. Yosys \cite{wolf2016yosys}  and ABC \cite{DBLP:conf/cav/BraytonM10} can be used to generate synthesizable Verilog code from any behavioral specification. NullaNet \cite{DBLP:conf/aspdac/NazemiPP19} generates the FFCL block in Verilog format and will be used as the upper stream engine. 
We first parse the Verilog netlist, synthesize the circuit using standard logic optimization techniques, primarily aimed at reducing the total gate count and depth of the circuit, and map the circuit to a customized cell library.
Notice that the Boolean operations supported by the logic gates in the cell library, such as two input AND, OR, and XOR operations, must be supported by the LPEs. 
Next, the mapped circuit is levelized \REVone{according to the definition in Section \ref{sec:notation}.} 
\REVone{The logic synthesis and levelization are the same as the one presented in \cite{arash_trets}.}

Because a gate that is at a specific logic level in a target circuit has no connections to any other gates at the same logic level,  operations of all gates at the same logic level can be executed simultaneously. However, these operations may have to be assigned to different compute cycles due to hardware resource limitations, i.e., the fixed number of LPVs per LPU (the \textit{depth issue})  or the fixed number of LPEs per LPV (the \textit{width issue}). Multiple LPUs can be assembled in parallel or series configuration for large graphs to complete the required computations for a given logic graph at the extra area/power cost. 
Our compiler has the ability to map any logic graph to an arbitrary-size LPU. To handle the width issue, the compiler decomposes the filter/neuron functions into MFGs, each of which is then mapped onto the LPU one after the other (more details in Section \ref{sec:compiler}).

\section{logic Processor Architecture}
\label{sec:Bool-proc-arc}

\begin{figure} [!tb]
    \centering
    \includegraphics[width=0.8\columnwidth]{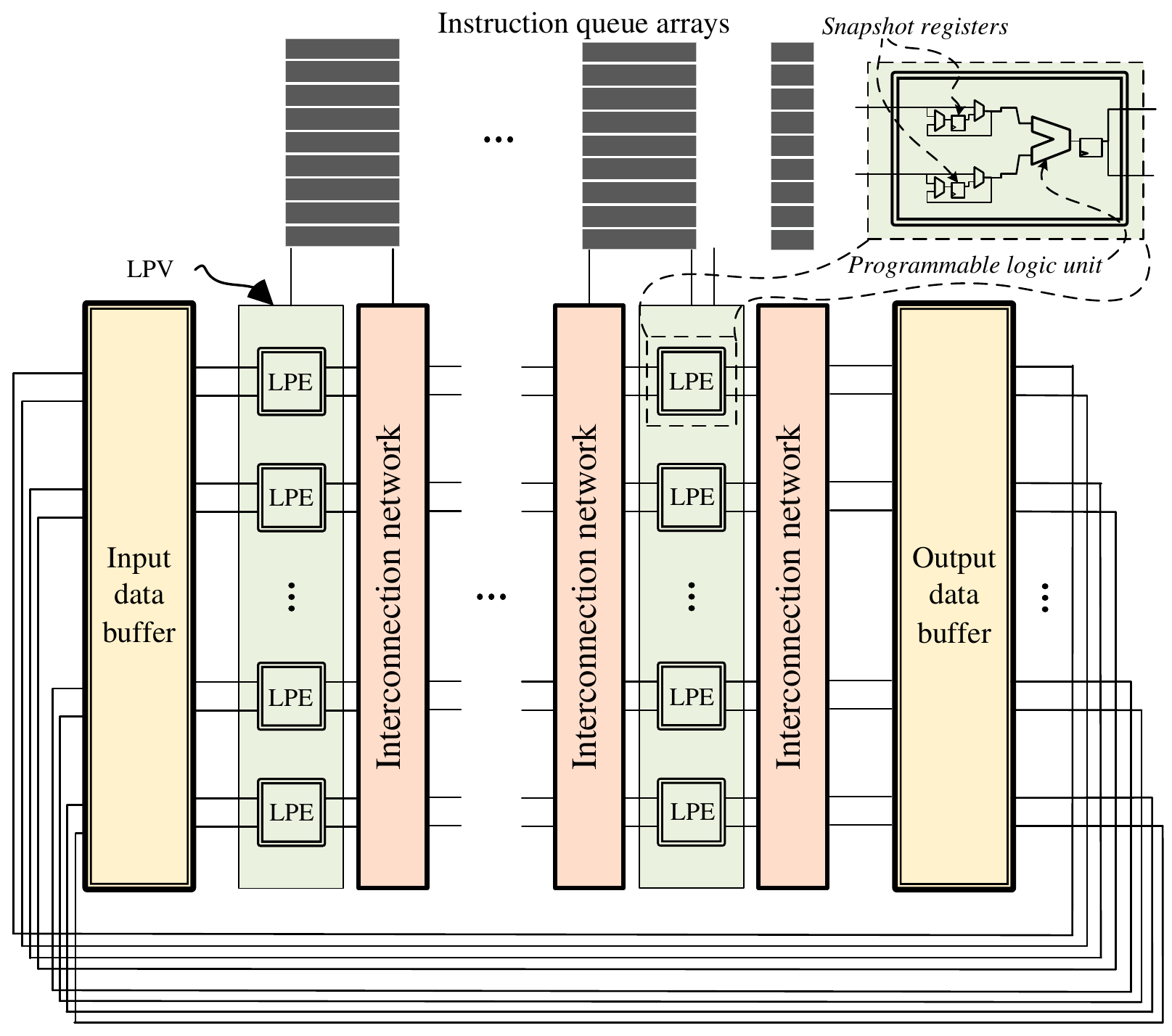}
    \caption{ \small The LPU architecture.}
    \label{fig:archi}
\vspace{-10pt}
\end{figure}

The architecture of the logic processor is shown in Fig. \ref{fig:archi}. The logic processor is a data-driven architecture in the sense that the streaming data is received and processed by multiple stages of \REVone{LPVs} without having to store any intermediate results in some scratchpad \REVone{memories}. More precisely, a logic processor comprises a set of \REVone{LPVs} that are linearly ordered. Each LPV contains $m$ \REVone{LPEs}, each of which receives two inputs and produces one output\DACREVone{, resembling a logic gate}. 
Therefore, each LPV receives up to $2m$ input operands and produces a vector of up to $m$ output results. To support logic operation packing and increase hardware efficiency, each operand has a width of $2m$ bits, which translates into $2m$ Boolean variables or $m$ 4-valued logic variables, etc. In the case of processing an FFCL block extracted from a convolutional neural network (CNN) model, the $2m$ bits of data come from different patches of an input feature volume in a CNN or from different 
images for batch-based inference tasks. To pass data from the $i$th LPV to the $(i+1)$th LPV, we use a non-blocking multicasting multi-stage switch network. 

\REVone{The number of LPEs per LPV and the number of LPVs per LPU determine the size of the logic graph that can be processed by an LPU}. With the parameter values described previously, an LPU can process a logic graph with a maximum width of $m$ and a maximum depth of $n$, where the width refers to the number of logic operations at any logic depth in the graph and the depth refers to the logic depth from any graph inputs to any graph outputs. 



Because of the data dependency between two adjacent logic levels within a logic graph, the intuitive way of computing a graph is level by level. However, given the hardware budget is tight and high throughput is desired, it is not an attractive option to use scratchpad memories to store temporary results (i.e., results produced but not yet consumed) between two logic levels. Therefore, we propose distributed \textit{snapshot registers} to store temporary results as shown in Fig. \ref{fig:archi}. Moreover, we propose a MFG-by-MFG \DACREVone{computing} paradigm and the scheduling algorithm, which are detailed in Section \ref{sec:compiler}. This paradigm further exploits the locality brought about by the \textit{snapshot registers} that are placed in each of the LPEs.

Each LPE contains a \textit{logic unit} where an elementary Boolean operation can be performed, and two \textit{snapshot registers} where each of the LPE inputs can be temporarily stored \DACREVone{for a certain data lifecycle determined by the compiler}. Additionally, the operations assigned to each LPE are configured with the aid of a instruction set. 
Notice two types of logic operation can be performed, namely, a multiple-input single-output (MISO) operation including AND, OR, XOR/XNOR and a single-input single-output (SISO) operation including NOT/BUFFER. BUFFER denotes a node that is added to a directed acyclic graph (DAG) to make all paths between any two connected nodes have the same topological length (i.e., an equal number of gates exist on all paths between two connected nodes). Buffer insertions are done as a part of the full path balancing step. \DACREVone{Full path balancing guarantees no data dependencies exist between two non-adjacent logic levels of gates, simplifying the mapping of the logic graph onto our pipelined architecture.}

\section{Compiler}\label{sec:compiler}

A compiler tailored to the logic processor is presented. The compiler parses a gate-level Verilog netlist to extract the set of operations that are carried out at each logic level of the circuit netlist, creates a DAG to represent these gate operations and their directional data dependencies, then partitions this DAG into MFGs such that each MFG can fit in the given LPU. \DACREVone{Then it schedules the generated MFGs and maps them to different LPVs in different compute cycles. As a result, the LPU processes level by level within each MFG and executes the large logic graph in an MFG-by-MFG manner.}

\subsection{Boolean network partitioning}

Consider a subgraph $H = (V', E')$ of a fully path balanced Boolean DAG $G = (V, E)$. Logic levels of nodes in subgraph $H$ are from $L_{bottom}(H)$ to $L_{top}(H)$. Now, given a partitioning solution $P$ of graph $G$ such that $P$ consists of $|P|$ subgraphs, we have:
\[\forall H\in P,~~~0\leq L_{bottom}(H)\leq L_{top} (H)\leq L_{max}\]
where
$0$ and $L_{max}$ denote the levels of primary inputs and primary outputs of $G$, respectively.
We denote the set of nodes in a logic level $l$ by $node\_set(l)$. Moreover, given a set of nodes $S$, let $input(S)$ denote the set of distinct nodes that feed into the set of nodes $S$. 

The MFGs must satisfy the following conditions.
First, the corresponding subgraphs $H$ must satisfy:
\begin{align}
\label{eqn: belongto}
    \forall l \in [L_{bottom}(H) + 1, L_{top}(H)] \implies \\ \nonumber input(node\_set(l)) \in H
\end{align}
which implies that inputs of all levels of a subgraph except the bottom-most level must also be contained in the same subgraph. Differently stated, inbound connections from nodes outside the subgraph to a node inside the subgraph are allowed only at the bottom-most level of the subgraph. 
Second, a subgraph must contain at most $m$ nodes in each level of the subgraph (recall that each LPV in the proposed logic processor has $m$ LPEs):
\begin{equation}
    \forall l \in [L_{bottom}(H), L_{top}(H)], |node\_set(l)|\leq m
\end{equation}
Note that nodes need not belong exclusively to a subgraph, i.e., MFGs can have overlapping node sets. More precisely,
\begin{equation}
    \exists l\in H, \exists l'\in H', node\_set(l) \cap node\_set(l') \neq \emptyset
\end{equation}
And, finally, nodes in the bottom-most level of an \DACREVone{MFG must} have input node count greater than $m$ with the exception of the subgraphs whose inputs are the PIs of $G$:
\begin{equation}
\begin{aligned} 
\forall H \in  P\setminus\{H_0|L_{bottom}(H_0)=0\}, \\
|input(node\_set(L_{bottom}(H)))| > m
\end{aligned}
\end{equation}
where $\setminus$ denotes the set subtraction operation. Notice that the inputs of the bottom-most level of a subgraph $H$ are not included in $H$. This equation is the break condition of while loop in Algorithm \ref{alg:InnerBFS}. 

\begin{figure} [!tb]
\vspace{-15pt}
    \centering
    \includegraphics[width=0.8\columnwidth]{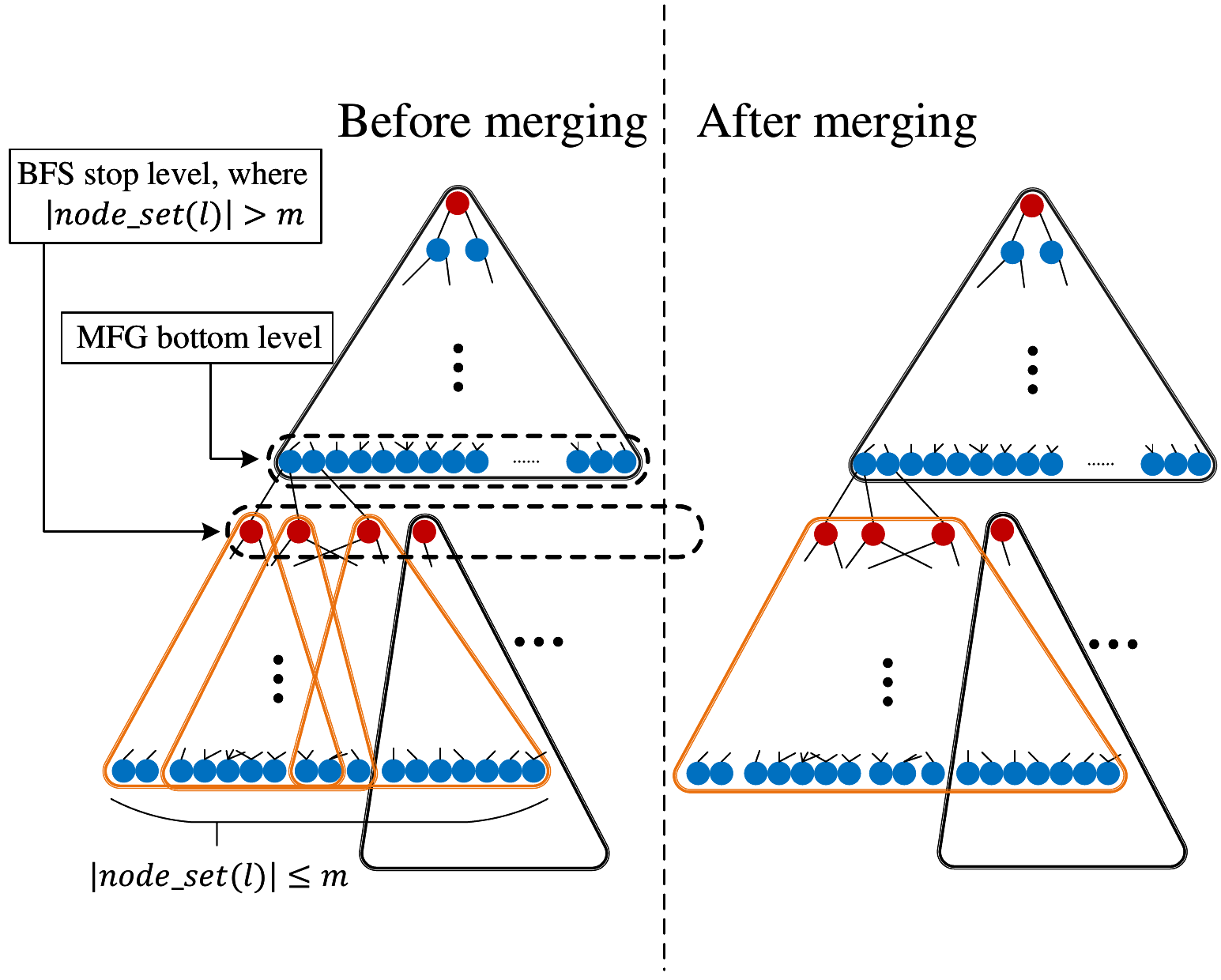}
    \caption{ \small An example of the merging procedure.}
    \label{fig:algo_explain}
\end{figure}

The Boolean network partitioning pseudo-code is shown in Algorithm \ref{alg:OuterBFS}. The algorithm uses a BFS traversal starting from the primary outputs (POs) on the given graph to find all MFGs.  
The MFG rooted at the POs is obtained by using the $findMFG()$ procedure as explained in Algorithm \ref{alg:InnerBFS}. Algorithm \ref{alg:OuterBFS} then continues by finding MFGs rooted at input nodes of the just extracted MFG. The traversal continues until we reach the PIs of the Boolean network. The procedure to construct an MFG rooted at some node $V$ keeps adding nodes to the MFG (again in a BFS traversal manner) until it reaches a logic level in its transitive fanin cone that has more than $m$ nodes, which is called the stop level. Note that, as seen in Algorithm \ref{alg:InnerBFS} and also shown in Fig. \ref{fig:algo_explain}, the MFG rooted at $V$ does not include the stop level nodes. The MFG cannot include any subset of nodes at its stop level because \REVone{condition} \ref{eqn: belongto} is violated in that case.

Algorithm \ref{alg:OuterBFS} returns a set of MFGs called allTempMFGs which contains 
MFGs that have one node at their top level, as shown in Fig. \ref{fig:algo_explain}. The runtime of a BNN inference task is primarily affected by the total number of MFGs. Therefore, a greedy merging algorithm (see Algorithm \ref{alg:merging}) is proposed to merge within a set of single-output MFGs that feeds into the same MFG and has the same bottom level, generates one multiple-output MFG (refer to Fig. \ref{fig:algo_explain}).
The checkLevel function checks whether the nodes of two MFGs in each logic level meet the constraint of $|$nodes(MFG1, $l_i$) $\cup$ nodes(MFG2, $l_i$)$|$ $\leq$ \textit{m}.
Note that one cannot merge two MFGs that have different bottom levels because of violating the property described in \REVone{condition} \ref{eqn: belongto}. Next, we discuss the main challenge we faced in compiler design. 

%

\begin{algorithm}[tb]
    \scriptsize
    \caption{\small Single-output Boolean network partitioning}
    \label{alg:OuterBFS}
    \begin{flushleft}
    \textbf{Input}: The PO of a Boolean network, $m$ number of LPEs per LPV\\
    \textbf{Output}: A set of MFGs that covers the Boolean network
    \end{flushleft}
    \begin{algorithmic}[1]
    \STATE allTempMFGs = $[ ]$   \COMMENT{a set of all MFGs}
    \STATE MFG=\textit{findMFG}(PO,$m$) \COMMENT{call Alg. \ref{alg:InnerBFS}}
    \STATE queue = $[ ]$
    \STATE queue.append(MFG)
    \WHILE{queue is not empty}
        \STATE curMFG = queue.pop(0)  \COMMENT{pop the first element} 
        \STATE allTempMFGs.append(curMFG)
        \FOR{inputNode in input(curMFG)}
            \STATE 
            childMFG=\textit{findMFG}(inputNode,$m$) \COMMENT{call Alg. \ref{alg:InnerBFS}}
            \STATE queue.append(childMFG)
        \ENDFOR
    \ENDWHILE
    \STATE \textbf{return} allTempMFGs
\end{algorithmic}
\end{algorithm}

\begin{algorithm}[tb]
    \scriptsize
    \caption{\small Function \textit{findMFG()}}
    \label{alg:InnerBFS}
    \begin{flushleft}
    \textbf{Input}: Node $V$, $m$ number of LPEs per LPV\\
    \textbf{Output}: A Graph object maintaining a MFG rooted at $V$
    \end{flushleft}
    \begin{algorithmic}[1]  
    \STATE MFG = Graph(root=V) \COMMENT{initialize a Graph with root=V}
    \STATE queue = $[]$
    \STATE queue.append($V$)
    \STATE visited = emptyset()
    \STATE visited.add($V$)
    \WHILE{queue is not empty}
        \STATE curNode = queue.pop(0)
        \STATE MFG.nodeCount[curNode.level]+=1
        \STATE visited.add(curNode)
        \IF{$|$MFG.nodeCount[curNode.level$|$ $\ge$ $m$} 
            \STATE stopLevel = curNode.level
            \STATE break
        \ENDIF
        \FOR{\REVone{child} in curNode.\REVone{children}}
            \IF{\REVone{child} not in visited}
                \STATE visited.add(\REVone{child})
                \STATE queue.append(\REVone{child})
            \ENDIF
        \ENDFOR
    \ENDWHILE
    \STATE MFG.bottomLevel = stopLevel + 1
    \STATE \textbf{return} MFG
\end{algorithmic}
\end{algorithm}

\subsection{Addressing the width issue}

For most of the Boolean networks generated by NullaNet \cite{DBLP:conf/aspdac/NazemiPP19}, the mapped netlists have sizes exceeding $n$ logic levels and contain a lot more than $m$ nodes per logic level. We refer to the former problem as the depth issue and the latter problem as the width issue, which is addressed \DACREVone{by the MFG-by-MFG computing paradigm as follows.}

MFGs can be scheduled and processed sequentially with the aid of the \textit{snapshot registers} to store the intermediate results generated by $node\_set(L_{top})$ of each MFG. In a pipelined manner, each MFG requires $L_{top}-L_{bottom}+1$ number of LPVs for its computation. Precisely, the computational resources allocated to MFG $H$ are $LPV \in[L_{bottom},L_{top}]$. The compute cycle count is $(L_{top}-L_{bottom}+1)\times t_c$ clock cycles where $t_c$ is the summation of one cycle for computation within an LPE and $t_{sw}$ cycles for data routing (steering) in the deployed switch network. \DACREVone{Note that $t_c$ is fixed for different BNN inference tasks. In this paper, $t_c=6$ because $t_{sw}=5$ where a 5-stage non-blocking multicast switch network \cite{yang1991nonblocking} is used.} The \textit{snapshot registers} of LPV associated with the $L_{top}+1$ of the MFG (e.g., $H_i$) to be routed, store the intermediate results generated by the given MFG. The parent MFG (i.e., $H_j$, the MFG that has inputs from the $H_i$) starts its computation at $L_{top}+1$ using the data stored in the \textit{snapshot registers}.

\begin{algorithm}[tb!]
    \scriptsize
    \caption{\small MFG merging algorithm}
    \label{alg:merging}
    \begin{flushleft}
    \textbf{Input}: A set of MFGs\\
    \textbf{Output}: Reduced set of MFGs
    \end{flushleft}
    \begin{algorithmic}[1]
    \STATE allMergedMFGs = $[ ]$
    \STATE rootMFG = the MFG contained PO(s)  
    \STATE queue = $[ ]$
    \STATE queue.append(topMFG)
    \WHILE{queue is not empty}
        \STATE curMFG = queue.pop(0)  \COMMENT{pop the first element} 
        \STATE allMergedMFGs.append(curMFG)
        \FORALL{MFG1 and MFG2 in curMFG.children}
             \STATE merge = False
            \IF{MFG1.bottomLevel == MFG2.bottomLevel}
                 \STATE merge = checkLevel(MFG1, MFG2)
            \ENDIF
            \IF{merge}
                \STATE mergedChildMFG $ = $ MFG1 $\cup$ MFG2
                \STATE queue.append(mergedChildMFG)
                \FORALL{grandchild in MFG1.children and MFG2.children}
                    \STATE update grandchild.parent to mergedChildMFG
                \ENDFOR
            \ELSE
                \STATE queue.append(MFG1, MFG2)
            \ENDIF
        \ENDFOR
    \ENDWHILE
    \STATE \textbf{return} allMergedMFGs
\end{algorithmic}
\end{algorithm}

\begin{figure} [!tb]
    \centering
    \includegraphics[width=0.4\columnwidth]{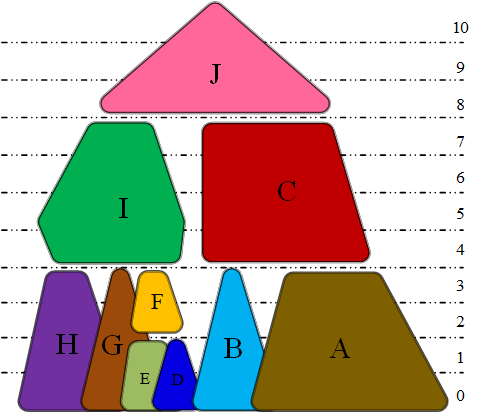}
    \caption{ \small An example of partitioning where $L_{max}=10$.}
    \label{fig:partition}
    \vspace{-8pt}
\end{figure}

\begin{figure} [!tb]
    \centering
    \includegraphics[width=0.8\columnwidth]{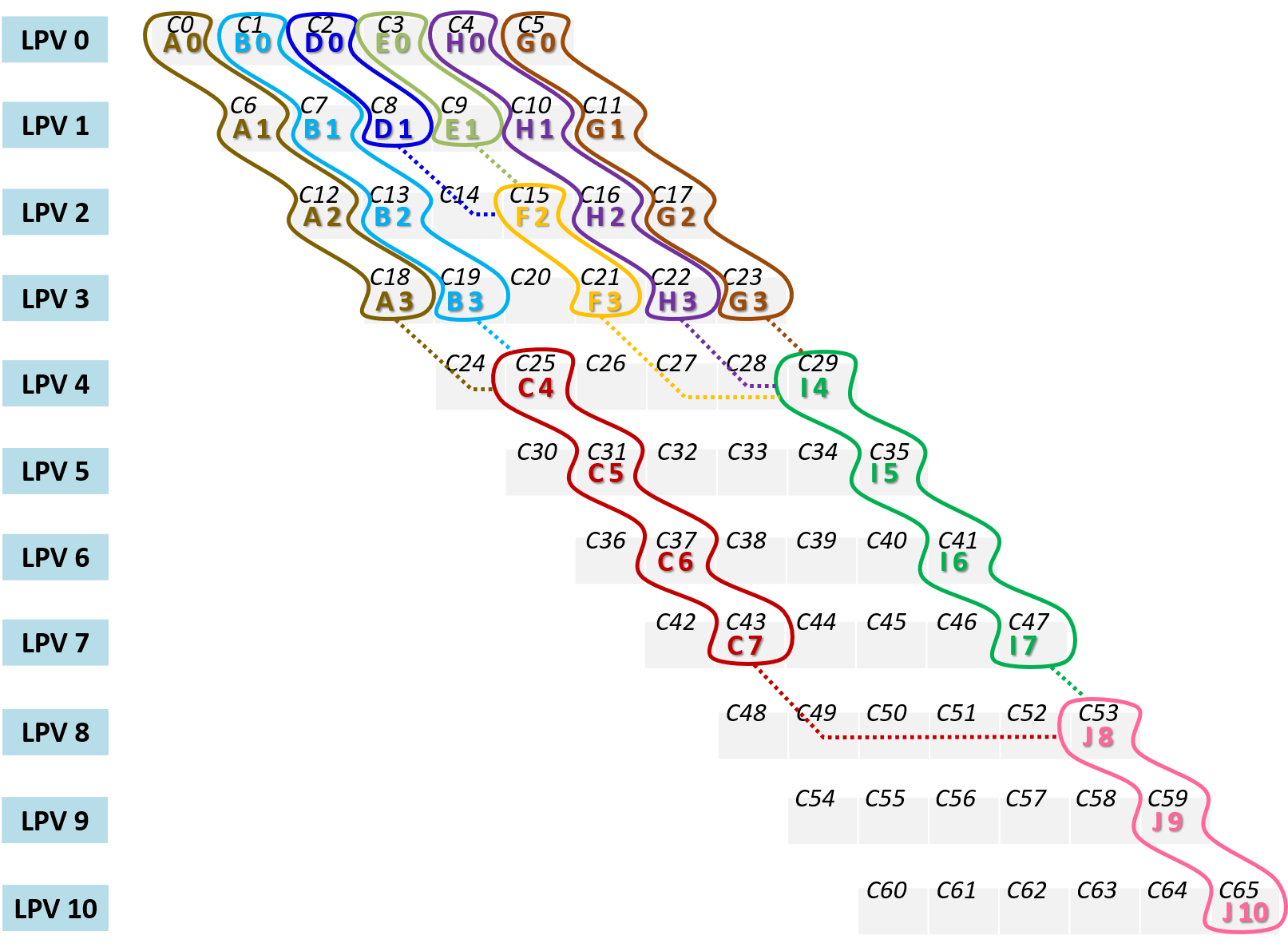}
    \caption{ \small  An example of scheduling. (Note: $C0$ denotes clock cycle 0)}
    \label{fig:schedule}
\vspace{-14pt}
\end{figure}

An illustrative graph partitioning solution is shown in Fig. \ref{fig:partition}. In this example, a Boolean network is partitioned into 10 MFGs, \REVone{which are shown with different colors}, each is scheduled and processed as shown in the time-space diagram of Fig. \ref{fig:schedule}. The computation periods of different logic levels are shown by ${l_i}$ after the MFG's name. For instance, $A1$ represents the computation of all nodes in logic level 1 of MFG $A$ and it takes 1 cycle. They are also associated with other two variables, computing cycles (i.e., C $j$) and LPVs (LPV $k$) to represent the computation of ${l_i}$th logic level of an MFG is performed in LPV $k$ at clock cycle $j$.

\begin{figure} [!t]

    \centering
    \includegraphics[width=0.85\columnwidth]{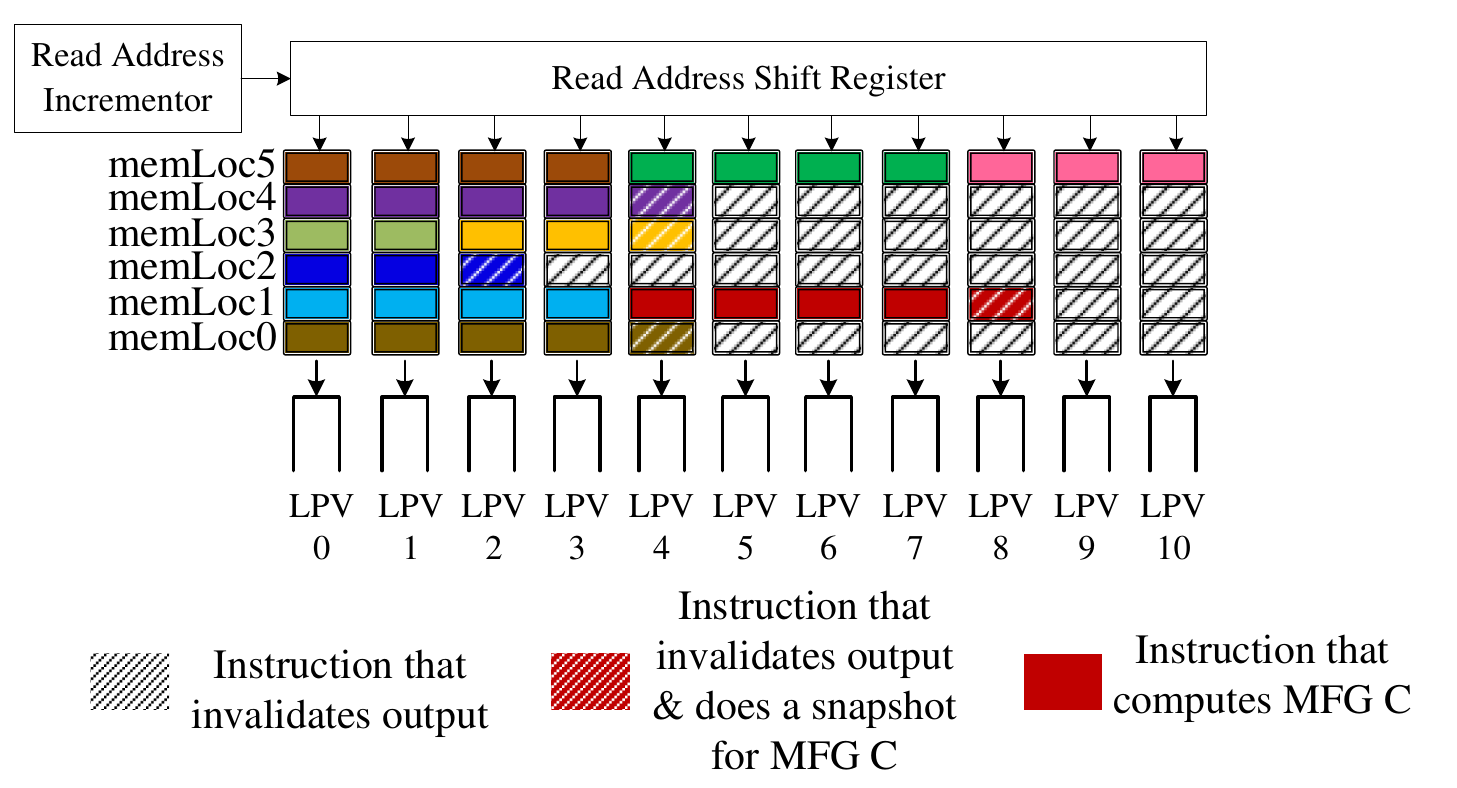}
    \vspace{-5pt}
    \caption{ \small Instruction queue configuration.}
    \label{fig:memLoc}
    \vspace{-3pt}
\end{figure}

In the LPU, a LPV stage and the 5 stages of the subsequent switch network form a block configured by a 6 instruction queues block, in which each memory takes the read address from its predecessor every cycle.
The instruction queues are accessible through a read address shift register.
The instructions corresponding to a MFG are written into the same address in all instruction queues associated with all LPVs that are involved in the computations of the MFG. For instance, instructions for computing MFG $J$, highlighted in pink in Fig. \ref{fig:schedule}, are written to memLoc5 of the instruction queues associated with LPVs \#8, 9, 10 as shown in Fig. \ref{fig:memLoc}.


In Fig. \ref{fig:memLoc}, memory locations of instruction queues, which program all the LPVs, are shown. The color of a memory location corresponds to the color of the MFG (cf. Fig. \ref{fig:partition}).
All MFGs with $L_{bottom}=0$ receive the PI values needed by $node\_set(L_{bottom})$ from the input data buffer. Using a counter, the compiler ensures that the required PI values are properly stored in different locations of the input data buffers such that the desired data is accessed correctly every cycle. 
This scheme \REVone{simplifies} the address generation compared to a random-access addressing system. Inputs of other MFGs that have $L_{bottom} \neq 0$, are provided by at least two child MFGs that have been computed earlier. Therefore, scheduling the MFGs is equivalent to arranging the instructions and placing them into proper memory locations.

\begin{algorithm}[!tb]
    \scriptsize
    \caption{\small MFG scheduling algorithm}
    \label{alg:Scheduling}
    \begin{flushleft}
    \textbf{Input}: A set of routed MFGs\\
    \textbf{Output}: Memory location to which instructions write
    \end{flushleft}
    \begin{algorithmic}[1]
    \STATE memLoc = $INT_{MAX}$
    \STATE topMFG.memLocation() = MemLoc
    \STATE stack = $[]$
    \STATE stack.append(topMFG)
    \WHILE{stack not empty}
    \STATE curMFG = stack.pop()
    \STATE curMFG.memLocation() = memLoc
        \IF{curMFG.parents $\neq$ None}
            \FOR{\REVone{childMFG} in curMFG.\REVone{children}()}
                \STATE stack.append(\REVone{childMFG})
            \ENDFOR
        \ELSE
        \STATE memLoc = memLoc - 1
        \ENDIF
        \FORALL{MFG}
    \STATE MFG.memLocation() = MFG.memLocation() -           \\memLoc
    \ENDFOR
    \ENDWHILE
\end{algorithmic}
\end{algorithm}

The scheduling algorithm shown in algorithm \ref{alg:Scheduling} determines the memory locations for writing the instructions for each MFG.  
As shown in Fig. \ref{fig:schedule}, each of the MFGs that satisfies $L_{bottom}\neq0$ has at least two \REVone{child} MFGs supporting all the nodes in its $node\_set(L_{bottom})$. For MFG $H$, we refer to the \REVone{child} MFGs as $H_c=\{H_{c1}, H_{c2},...,H_{cn}\}$. Each of the MFGs in $H_c$ will generate a subset of $input(node\_set(L_{bottom}(H)))$; therefore, we have $\bigcup_{i=1}^{n} node\_set(L_{top}(H_{ci}))=input(node\_set(L_{bottom}(H)))$. We refer to the last-scheduled \REVone{child} MFG that generates a subset of the \REVone{parent's} input nodes (without having to store them in the snapshot registers) as the most recent \REVone{child}. For instance, MFG $J$'s most recent \REVone{child} is MFG $I$ (see Fig. \ref{fig:schedule}). Note that two MFGs may use the same memLoc value as long as one of them is the most recent \REVone{child} of the other (e.g., MFG $I$, $J$) because they are performed on different LPVs. So, the required size of the instruction queues is reduced.

\subsection{Addressing the depth issue}

The depth issue is left to the LPU hardware. If an MFG has $L_{top}$ that is greater than the total amount of LPVs \REVone{of a fixed size LPU}, the output data buffer will perform as the snapshot registers of LPV $L_{top}+1$, which is the LPV that does not physically exist. Instead, LPV 0 resorts to the circulation mechanism to perform the functionality of LPV $L_{top}+1$. The compiler is responsible for detecting potential depth issue and relocating the instructions accordingly. The compiler notifies the hardware when to feed the intermediate data stored in the output data buffer back to the LPV 0. Such data goes through the pipeline for multiple rounds of computation to complete the computation of all logic levels of the given FFCL. 

\section{Experimental Results}
\label{sec:results}
For evaluation purposes, we targeted a high-end Virtex\textregistered ~UltraScale+ FPGA (Xilinx VU9P FPGA 
, which is available in the cloud as the AWS EC2 F1 instance). We include the FPGA prototyping results since the \REVone{SoA} implementations use the same FPGA.
The hardware metrics are reported in table \ref{table:res-uti} for LPV count = 16.

\REVone{Our benchmarked models can be categorized into two groups, i) models for high-accuracy requirement (i.e., large models), and ii) models for high-throughput requirement (i.e., tiny models). 
In the first group, we study VGG-16, VGG-like model used in ChewBaccaNN\cite{DBLP:conf/iscas/AndriKCB21}, LENET5 on the MNIST dataset, and MLPMixers \cite{DBLP:conf/nips/TolstikhinHKBZU21} on the CIFAR-10 dataset.}
\REVone{We also evaluate our logic processor against extreme-throughput tasks in physics and cybersecurity such as jet substructure classification (JSC) \cite{DBLP:journals/corr/abs-1804-06913} and network intrusion detection (NID) \cite{article-BNN}. We used UNSWNB15 dataset to compare our logic processor with other implementations. We employed the same preprocessed training and testing data as that of Murovic et al. \cite{article-BNN} which has 593 binary features corresponding to 49 original features and two output classes. }

For MLPMixers, the resolution of the input image is 32*32, and the patch size that the experiments are based on is 4*4. So, we have 64 non-overlapping image patches that are mapped to a hidden dimension $C$ which is 128 and 192 for small design (S) and Base design (B), respectively. $D_{S}$ and $D_{C}$ are tunable hidden widths in the token-mixing and channel-mixing MLPs, respectively. $D_{S}$ and $D_{C}$ are 64 (96) and 512 (768) for S (B) design. There are 8 and 12 mixing layers in S and B designs, respectively. 




\begin{table}[tb]
\vspace{-10pt}
  \caption{ \small Resource utilization of design of LPV count = 16.} 
  \label{table:res-uti}
  \scriptsize
  \centering 
  \renewcommand{\arraystretch}{0.8}
  \begin{tabular}{|c|c|c|c|} 

  \midrule
  
  FF(\%) & LUT(\%) & BRAM(\%) & FREQ\\
  \midrule
  478K(20.2\%) & 433K(36.7\%) & 12240K (15.8\%) & 333MHz \\
  \midrule

  \end{tabular}


\end{table}

\subsection{Effect of the MFG merging procedure}
\label{subsec:comp-opt-res}
\begin{figure} [tb!]
\vspace{-10pt}
\centering
    \begin{subfigure}{.5\columnwidth}
      \centering
      \includegraphics[width=\textwidth]{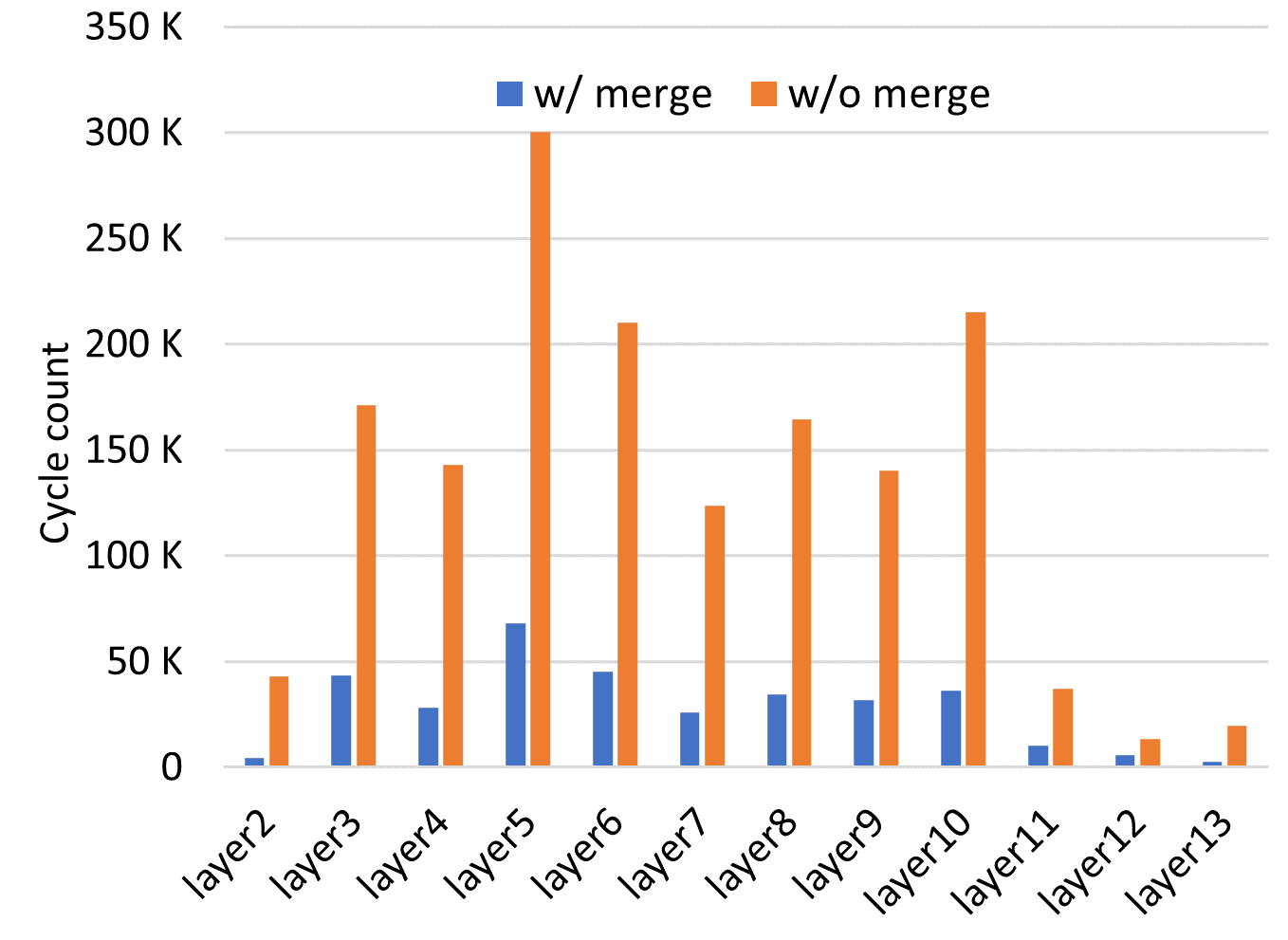}
      \caption{Computation time}
      \label{fig:sub1}
    \end{subfigure}%
    \begin{subfigure}{.5\columnwidth}
      \centering
      \includegraphics[width=\textwidth]{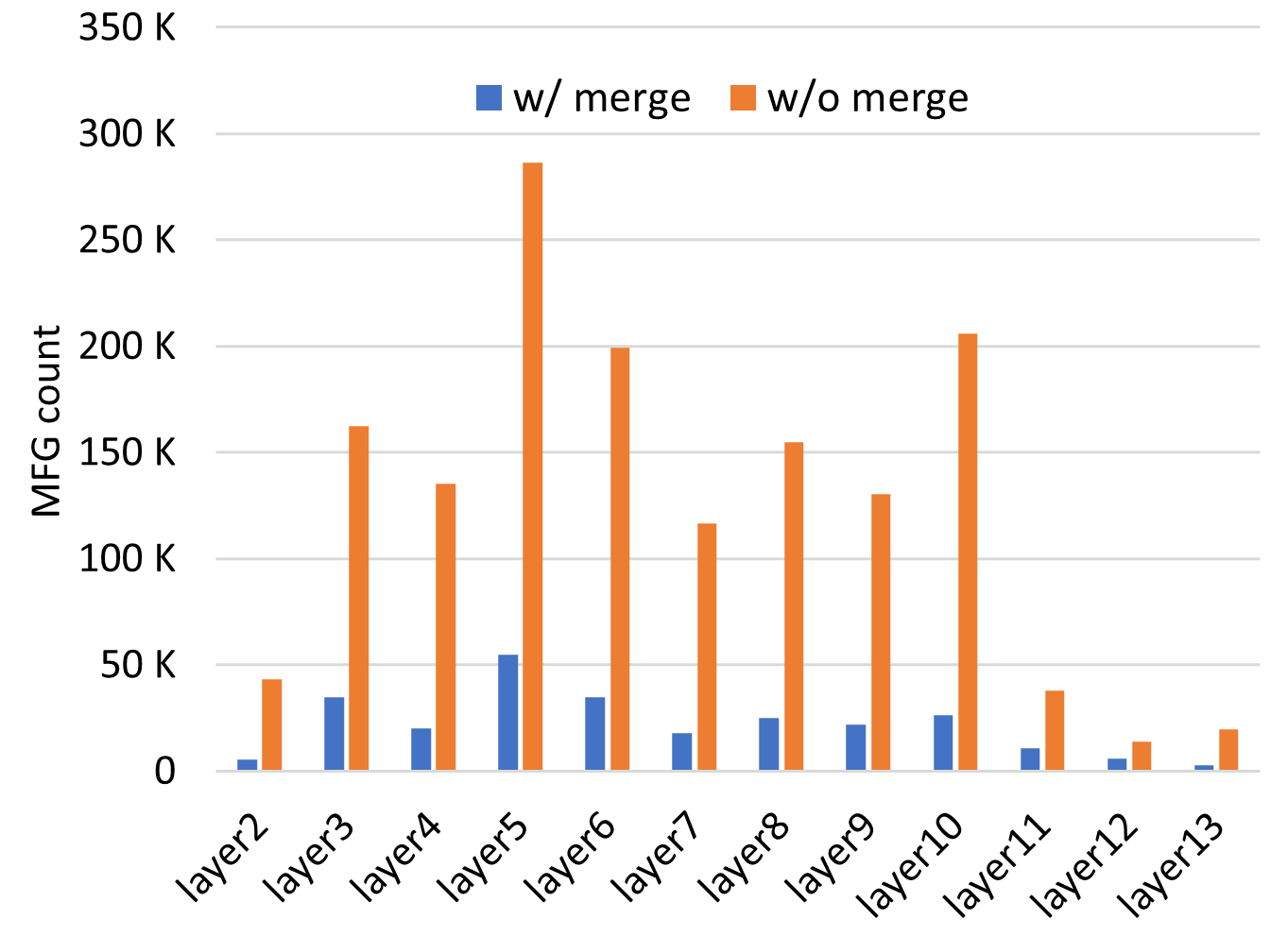}
      \caption{MFG count}
      \label{fig:sub2}
    \end{subfigure}
\caption{\small \REVone{a) Computation time and b) MFG count of VGG16 layers before and after applying merging algorithm.}}
\label{fig:withoutmerge}
\end{figure}

\begin{figure} [!tb]
\vspace{-20pt}
\centering
    \begin{subfigure}{.5\columnwidth}
      \centering
      \includegraphics[width=\textwidth]{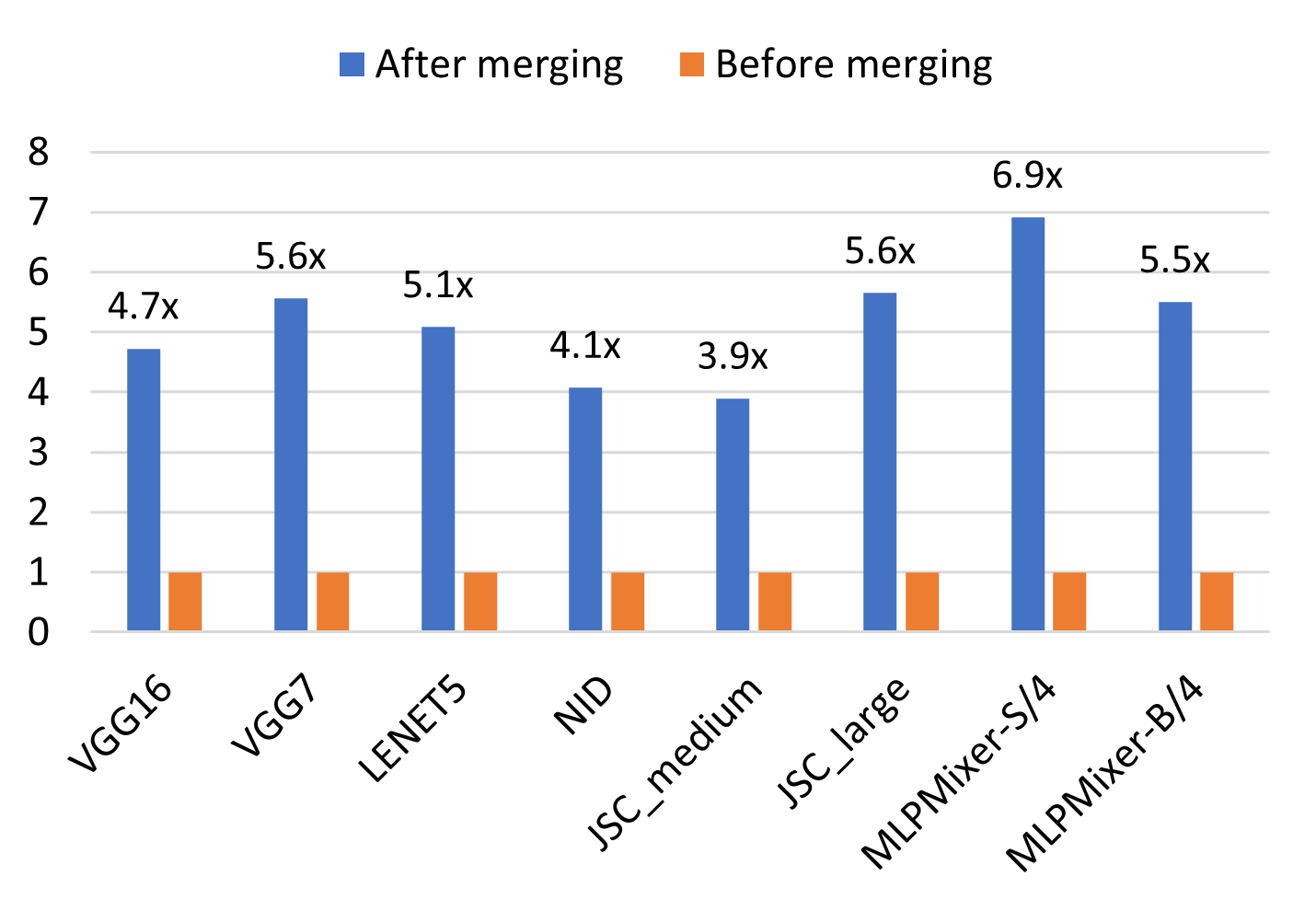}
      \caption{Throughput}
      \label{fig:models_th}
    \end{subfigure}%
    \begin{subfigure}{.5\columnwidth}
      \centering
      \includegraphics[width=\textwidth]{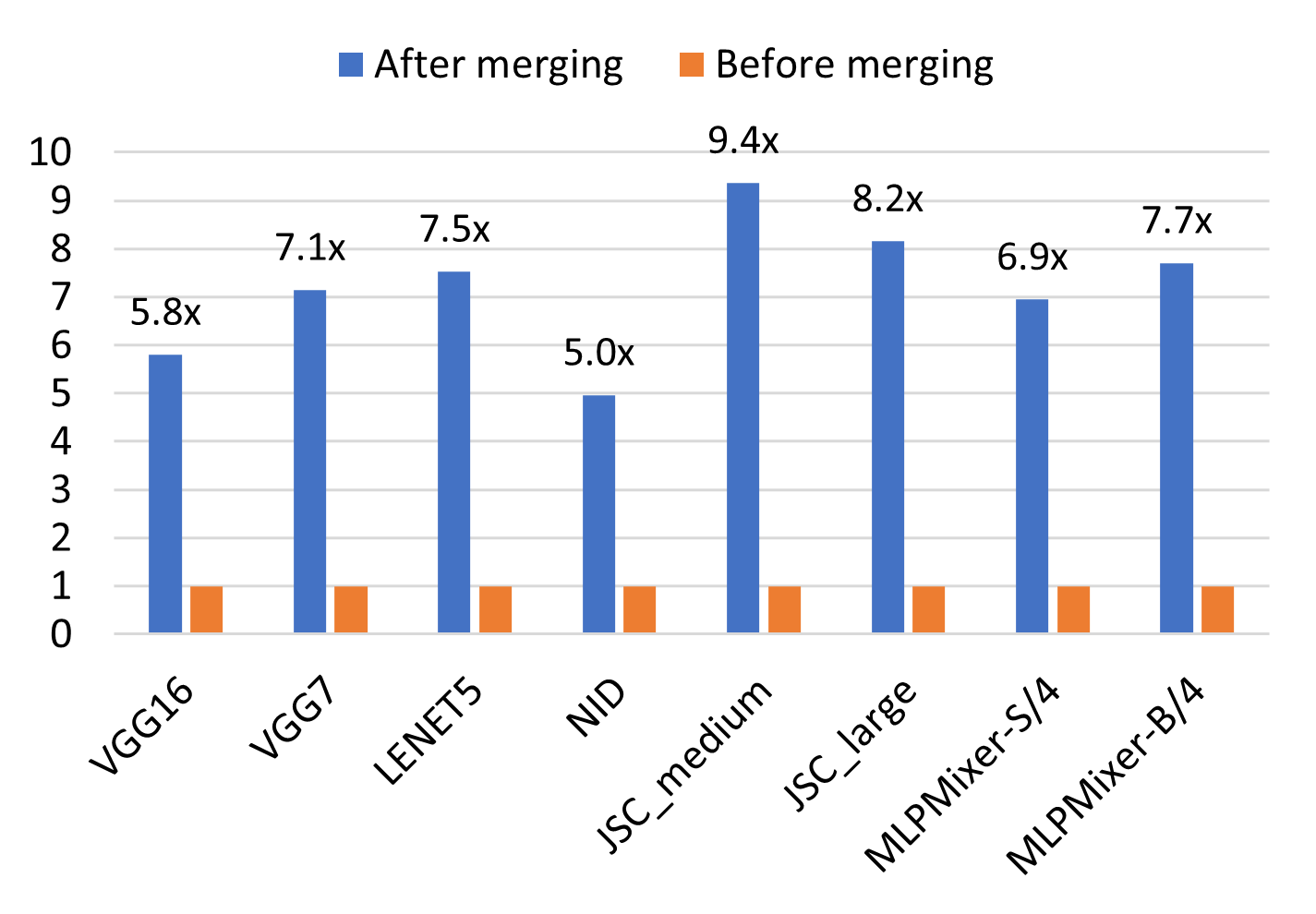}
      \caption{MFG count}
      \label{fig:models_mfg}
    \end{subfigure}
\caption{\small \REVone{a) Throughput and b) MFG count before and after applying merging algorithm.}}
\label{fig:merge_vs_unmerge_different_models}
\end{figure}

To show the efficacy of the proposed merging procedure described in Section \ref{sec:compiler}, we compare the performance of our logic processor with and without incorporating the merging procedure. \REVone{Results are shown in Fig. \ref{fig:withoutmerge} and \ref{fig:merge_vs_unmerge_different_models}. Fig. \ref{fig:sub1} shows the clock cycle count for computing layers [2:13] of VGG16 with and without the merging procedure. Fig. \ref{fig:sub2} shows the total number of MFGs obtained from the proposed algorithm with and without the merging procedure. The results confirm the superior performance after applying the merging procedure and also demonstrate the high correlation between computation time and the MFG count. To further assess the effectiveness of the proposed merging algorithm, we apply it to all models used in this study and summarize the results in Fig. \ref{fig:merge_vs_unmerge_different_models}. As can be seen, the throughput is improved by 5.2x on average while the MFG count can be reduced up to 9.4x. }

\subsection{Comparison Between MAC-based, XNOR-based and NullaNet-based FPGA Implementation of NNs}
\label{subsec:comparisons}
The VGG16 is a huge network and it has about 138 million parameters. We implement intermediate convolutional layers 2-13 in VGG16 using the proposed framework and fixed-function combinational logic functions. 
As a baseline for the SoA generic MAC array-based accelerator for the layers realized using conventional MAC calculations, we used the open-source implementation of \cite{DBLP:conf/fpga/SohrabizadehWC20} with some improvements proposed in \cite{arash_trets}. 
We use FINN \cite{DBLP:conf/fpga/UmurogluFGBLJV17} for our XNOR-based baseline. We improve this implementation by packing operations. We also use the NullaDSP model \cite{arash_trets} as another baseline for mapping FFCL generated by NullaNet to the programmable DSP blocks where it can fit any FFCL with any size. We use the best results of each implementation reported in \cite{arash_trets} and compare them to our proposed architecture in tables \ref{tab:FPS_high_acc} and \ref{tab:FPS_high_th}.  \REVone{In the case of JSC and NID, we use the implementation and the associated performance reported in LogicNets \cite{DBLP:conf/fpl/UmurogluAFB20}, Google and CERN's optimized implementation \cite{DBLP:journals/natmi/CoelhoKLZNALPPS21}}, and the implementation presented in \cite{DBLP:journals/corr/abs-2201-11409}. 

As illustrated in the tables \ref{tab:FPS_high_acc}, our \REVone{implementation} shows superior performance compared to other implementations. The LPU and XNOR implementation achieves significant saving, especially in large DNN model like VGG16, since we keep all intermediate data on on-chip memories and there is no cost \REVone{associated} with off-chip memories while this is not the case for MAC-based and NullaDSP implementation.
We achieved 14.01x(33.43x), 4.86x(3.93x), and 1.95x(4.89x) in performance improvement on VGG16 (LENET-5) inference compared to MAC-based, NullaDSP, XNOR-based implementations, respectively. 

\REVone{LogicNets \cite{DBLP:conf/fpl/UmurogluAFB20} have higher frames per second (FPS) than our design. However, they cannot use the same hardware for the other models since they realized each model as a customized hard network of logic gates (as in random logic blocks). Whereas, our design offers programmable logic processors that can perform the required logic gate operations of any logic (computation) graphs. The former realization is ideal for building a highly efficient, yet unchangeable, inference engine whereas the latter one is desirable for accelerating the training process and for building inference engines that can be updated after they are deployed in the field. Note that NullaNet Tiny \cite{DBLP:conf/fccm/NazemiFEKSP21} is our upstream for generating FFCL blocks and presents a similar implementation as LogicNets \cite{DBLP:conf/fpl/UmurogluAFB20} and outperforms the LogicNets in similar settings on the same benchmarks.}
%


\begin{table}[!t]
  \caption{ \small\REVone{FPS Comparison between the different implementation of models with high accuracy requirements. LPV count in LPU is 16.}} 
  \label{tab:FPS_high_acc}
  \scriptsize
  \centering 
  \renewcommand{\arraystretch}{0.8}
  \begin{tabular}{|l|c|c|c|c|} 
  \midrule
  
   & MAC & NullaDSP & XNOR & LPU \\
  \midrule
  VGG16 & 0.12K & 0.33K & 0.83K & \textbf{103.99K} \\
  \midrule
  LENET5 & 0.48K & 4.12K & 3.31K & \textbf{1035.60K} \\
  \midrule
  MLPMixer-S/4 & 4.17K & - & 50.00K & \textbf{179.23K} \\
  \midrule
  MLPMixer-B/4 & 0.88K & - & 16.67K & \textbf{102.01K} \\
  \midrule
  \end{tabular}



\end{table}

\begin{table}[tb]
  \caption{ \small\REVone{FPS Comparison between the different implementation of models with high throughput requirements. LPV count in LPU is 16.}} 
  \label{tab:FPS_high_th}
  \scriptsize
  \centering 
  \renewcommand{\arraystretch}{0.8}
  \begin{tabular}{|l|c|c|c|c|} 
  \midrule
  
   & LogicNets \cite{DBLP:conf/fpl/UmurogluAFB20} & Google+CERN\cite{DBLP:journals/natmi/CoelhoKLZNALPPS21} & \cite{DBLP:journals/corr/abs-2201-11409} & LPU \\
  \midrule
  NID & \textbf{95.24M} & - & 49.58M & 8.39M \\
  \midrule
   JSC-M & \textbf{2995.00M} & - & - & 0.69M \\
   \midrule
  JSC-L & \textbf{76.92M} & 76.92M & - & 0.21M \\
  \midrule
  \end{tabular}


\vspace{-12pt}
\end{table}


\begin{figure} [!tb]
\vspace{-5pt}
    \centering
    \includegraphics[width=\columnwidth]{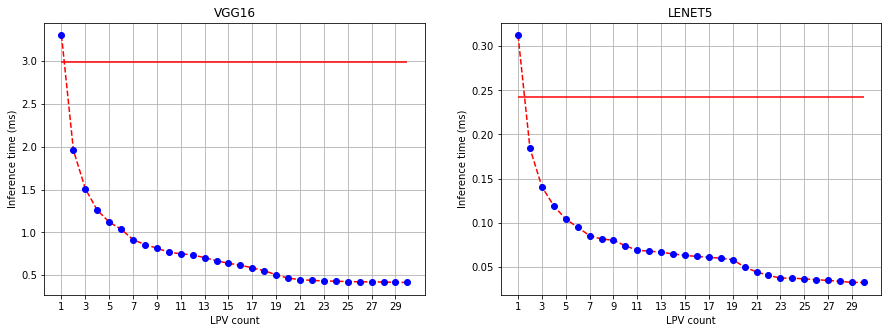}
    \caption{ \small Inference time of VGG16 and LENET5.}
    \label{fig:LPV_alter_result}
\end{figure}

\subsection{Ablation study with LPV count}
\label{subsec:comparisons}
To determine the influence of the LPV count on the performance of the presented logic processor, we conducted experiments with different LPV counts. 
As shown in Fig. \ref{fig:LPV_alter_result}, the inference time decreases as we increase the number of LPVs. The influence of the LPV count is saturated after a while. 
To conduct a comparative analysis on NullaDSP \cite{arash_trets} against the LPU, we benchmark the effective LPV threshold, which is defined as the minimum number of LPVs that an LPU needs to achieve performance equivalent to NullaDSP. As shown in Fig. \ref{fig:LPV_alter_result}, we need at least \REVone{2} LPVs to achieve such performance for the case of VGG16.

\section{Conclusion}
\label{sec:conc}
The proposed logic processor offers a novel hardware design approach afforded by the presented compiler resulting in efficient partitioning and mapping of a given neural network model in the format of fixed-function combinational logic.
In future work, we plan to intend to explore the heterogeneous architecture where the number of LPEs per LPVs and their following switch networks will not be the same for all LPVs. Also, it is worth trying multiple LPUs that can be assembled in parallel or series configurations. 


{\small \bibliographystyle{IEEEtranS}
\bibliography{refs}}

\end{document}